\documentclass[a4paper]{article}
\usepackage{jheppub}
\usepackage{times}
\usepackage{amsmath}
\usepackage{amsfonts}
\usepackage{graphicx}
\usepackage{dcolumn}
\usepackage{bm}

\def\beq{\begin{equation}}
\def\eeq{\end{equation}}
\def\bea{\begin{eqnarray}}
\def\eea{\end{eqnarray}}

\def\ep{\epsilon}

\def\nn{\nonumber}
\def\Eq#1{Eq.~(\ref{#1})}
\def\ln#1{\mathrm{log}\left(#1\right)}

\def\td#1{\tilde{\delta}\left(#1\right)}

\def\cgt{\widetilde{c}_{\Gamma}}
\newcommand{\la}{\langle}
\newcommand{\ra}{\rangle}

\def\qb{\mathbf{q}}

\def\uv{{\rm UV}}
\def\cnt{{\rm cnt}}
\def\r{{\rm R}}
\def\v{{\rm V}}
\def\f{{\rm f}}
\def\b{{\rm b}}

\begin{document}

\preprint{IFIC/15-35, LPN15-025}

\title{Towards gauge theories in four dimensions}

\author[a]{Roger J. Hern\'andez-Pinto}
\author[a,b]{, Germ\'an F. R. Sborlini}
\author[a]{and Germ\'an Rodrigo} 
\affiliation[a]{Instituto de F\'{\i}sica Corpuscular, Universitat de Val\`{e}ncia -- Consejo Superior de Investigaciones Cient\'{\i}ficas, 
Parc Cient\'{\i}fic, E-46980 Paterna, Valencia, Spain.}
\affiliation[b]{Departamento de F\'\i sica and IFIBA, FCEyN, Universidad de Buenos Aires, 
(1428) Pabell\'on 1 Ciudad Universitaria, Capital Federal, Argentina.}

\emailAdd{rogerjose.hernandez@ific.uv.es}
\emailAdd{german.sborlini@ific.uv.es}
\emailAdd{german.rodrigo@csic.es}


\date{August 18, 2015}

\abstract{
The abundance of infrared singularities in gauge theories 
due to unresolved emission of massless particles (soft and collinear) 
represents the main difficulty in perturbative calculations.  
They are typically regularized in dimensional regularization, 
and their subtraction is usually achieved independently for virtual and real corrections. 
In this paper, we introduce a new method based on the loop-tree duality (LTD) theorem 
to accomplish the summation over degenerate infrared states directly at the integrand level 
such that the cancellation of the infrared divergences is achieved simultaneously, and 
apply it to reference examples as a proof of concept. 
Ultraviolet divergences, which are the consequence of the point-like nature of the theory,
are also reinterpreted physically in this framework. The proposed method opens the intriguing 
possibility of carrying out purely four-dimensional implementations of higher-order 
perturbative calculations at next-to-leading order (NLO) and beyond free of soft and final-state 
collinear subtractions.}

\maketitle


\section{Introduction}
\label{sec:intro}

The most important Quantum Field Theories (QFT) for describing 
elementary particle physics are gauge theories. Quantum Yang-Mills theories~\cite{Yang:1954ek}
were introduced about sixty years ago and are
now the foundation of most of elementary particle theory. 
One of the main difficulties to extract theoretical predictions 
from perturbative calculations at higher orders is the spread of infinities. 
In the high-energy regime, there are ultraviolet (UV) divergences related to the point-like
nature of the theory. Due to the preservation of gauge invariance, 
dimensional regularization (DREG)~\cite{Bollini:1972ui,'tHooft:1972fi,Cicuta:1972jf,Ashmore:1972uj}, 
that consists in changing the number of space-time 
dimensions from $4$ to $d=4-2\ep$, is the customary regularization method to overcome
this problem. The UV singularities in four dimensions manifest as poles in the dimensional 
parameter $\ep$, and are easily removed by renormalization. The real challenge is, 
however, in the infrared (IR) regime.  Unresolved radiation of massless particles (soft and collinear) 
is physically degenerate with the absence of emission, 
while from the theory point of view they are described
as different final states. This leads to further infinities that are also 
regularized in DREG.

The standard 
approach~\cite{Catani:1996vz,Catani:1996jh,Frixione:1995ms,GehrmannDeRidder:2005cm,Catani:2007vq,DelDuca:2015zqa,Czakon:2010td,Boughezal:2015dva,Gaunt:2015pea} 
to remove the IR singularities consists of adding to 
the real radiation contribution suitable subtraction terms that mimic the IR
behavior of the emission scattering amplitudes and render them finite in the 
IR limit. The same subtraction quantities, integrated over the phase-space of the extra
radiation, are subtracted back from the virtual corrections.  The sum of both 
contributions leads to finite theoretical predictions for physical observables, 
if all the degenerate states are considered in the sum. In this paper,
we describe an alternative approach which is based on the loop-tree duality (LTD)
theorem~\cite{Catani:2008xa,Bierenbaum:2010cy,Bierenbaum:2012th,Buchta:2014dfa,Buchta:phd,Buchta:2015vha,Buchta:2015wna,Sborlini:2015uia}. 
We propose a new method that combines simultaneously 
virtual and real corrections, following in fact the original aim of the LTD method.
We present here the first practical realization of LTD that 
we illustrate with some reference examples. This fact has also a strong implication: the possibility of carrying out purely four-dimensional implementations free of soft 
and final-state collinear subtractions. 
Unlike the method proposed in Ref.~\cite{Soper:1998ye}, LTD does not require 
any smearing in the total energy to combine real and virtual corrections.  
Likewise, massless internal propagators are kept massless in LTD, 
as opposed to the four dimensional implementations presented in 
Refs.~\cite{Pittau:2012zd,Donati:2013iya,Fazio:2014xea}.

\section{Review of the loop-tree duality}
\label{sec:one-loop}

The LTD theorem~\cite{Catani:2008xa} establishes a direct dual relation 
between one-loop $N$-leg scalar integrals or scattering amplitudes 
in any relativistic, local and unitary QFT with corresponding tree-level objects 
integrated in a modified phase-space that resemble real corrections. 
Specifically, the dual representation of a $N$-leg scalar one-loop integral consists 
of the sum of $N$ dual integrals:
\bea
L^{(1)}(p_1, \dots, p_N) 
&=& - \sum_{i\in \alpha_1} \, \int_{\ell} \; \td{q_i} \,
\prod_{j \in \alpha_1, \, j\neq i} \,G_D(q_i;q_j)~, \nn \\
\label{oneloopduality}
\eea 
where $G_D(q_i;q_j) = (q_j^2 -m_j^2 - i0 \, \eta \cdot k_{ji})^{-1}$
are dual propagators, with $i,j \in \alpha_1 = \{1,2,\ldots N\}$. 
The momenta of the internal lines are denoted $q_{i,\mu} = (q_{i,0},\mathbf{q}_i)$, 
where $q_{i,0}$ is the energy (time component) and $\qb_{i}$ are 
the spacial components. They are defined as $q_{i} = \ell + k_i$ with 
$\ell$ the loop momentum and $k_{i} = p_{1} + \ldots + p_{i}$. 
The four-momenta of the external legs are $p_{i}$, which are taken as 
outgoing, with $k_{N} = 0$ by momentum conservation. 
The loop integration measure reads
\beq
\int_{\ell} \bullet 
= - i \mu^{4-d} \int \frac{d^d \ell}{(2\pi)^{d}} \bullet~.
\eeq
The delta function 
$\td{q_i} \equiv 2 \pi \, i \, \theta(q_{i,0}) \, \delta(q_i^2-m_i^2)$
sets the internal lines on-shell and is the result of selecting the poles of the Feynman 
propagators with positive energy $q_{i,0}$ and negative imaginary part by applying 
the Cauchy residue theorem along a contour that is closed in the lower half 
of the complex plane. LTD is equivalent to perform the loop 
integration in the $d-1$ dimensional space defined by the forward on-shell hyperboloids, 
$G_F(q_i)^{-1}=(q_i^2-m_i^2+i0)=0$ and $q_{i,0}>0$, or  
forward light-cones for massless propagators.

The key feature of the dual representation in~\Eq{oneloopduality}
is that the usual Feynman propagators have been replaced by dual propagators 
whose $+i0$ prescription depends now on the sign of $\eta \cdot k_{ji}$, 
where $\eta$ is a {\em future-like} vector,  $\eta^2 \ge 0$, with positive definite energy $\eta_0 > 0$,
and $k_{ji} = q_j - q_i$. The idea of having different $+i0$ prescriptions for different 
propagators is at first 
sight astonishing, but indeed it is a necessary condition for the consistency of the method. 
As shown in Ref.~\cite{Buchta:2014dfa}, the integrand in \Eq{oneloopduality} becomes 
singular at the intersection of forward on-shell hyperboloids 
(light-cones for massless propagators), and forward with backward ($q_{j,0}<0$) on-shell 
hyperboloids. Those singularities lead to integrable threshold singularities or 
non-integrable soft and collinear divergences. In the forward--forward case, there is a cancellation 
of singularities among different dual contributions, and the change of sign of the 
dual prescription plays a central role in that cancellation. 
In the forward--backward case, the singularities 
remain constrained to a compact region of the loop three-momentum and admit a nice physical 
interpretation in terms of causality, which is indeed consistent with the Cutkosky rule. 
These singularities occur only in one direction, the direction of the chosen 
internal momentum flow, when the on-shell virtual particle
is emitted and interacts with another outgoing virtual on-shell 
particle ($q_{j,0}<0$) after loosing energy by radiating external particles.

\section{Dual cancellations}
\label{sec:cancel}

\begin{figure}[ht]
\begin{center}
\includegraphics[width=10cm]{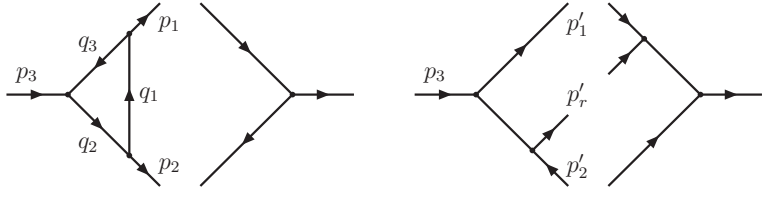}
\caption{Kinematic configuration of the three-point function for the process 
$p_3\to p_1+p_2$, its interference with the Born amplitude (left), and interference 
of real radiation contributions with $p_3 \to p_1'+p_2'+p_r'$ (right).  
\label{fig:feynman}}
\end{center}
\end{figure}

\begin{figure}[ht]
\begin{center}
\includegraphics[width=6cm]{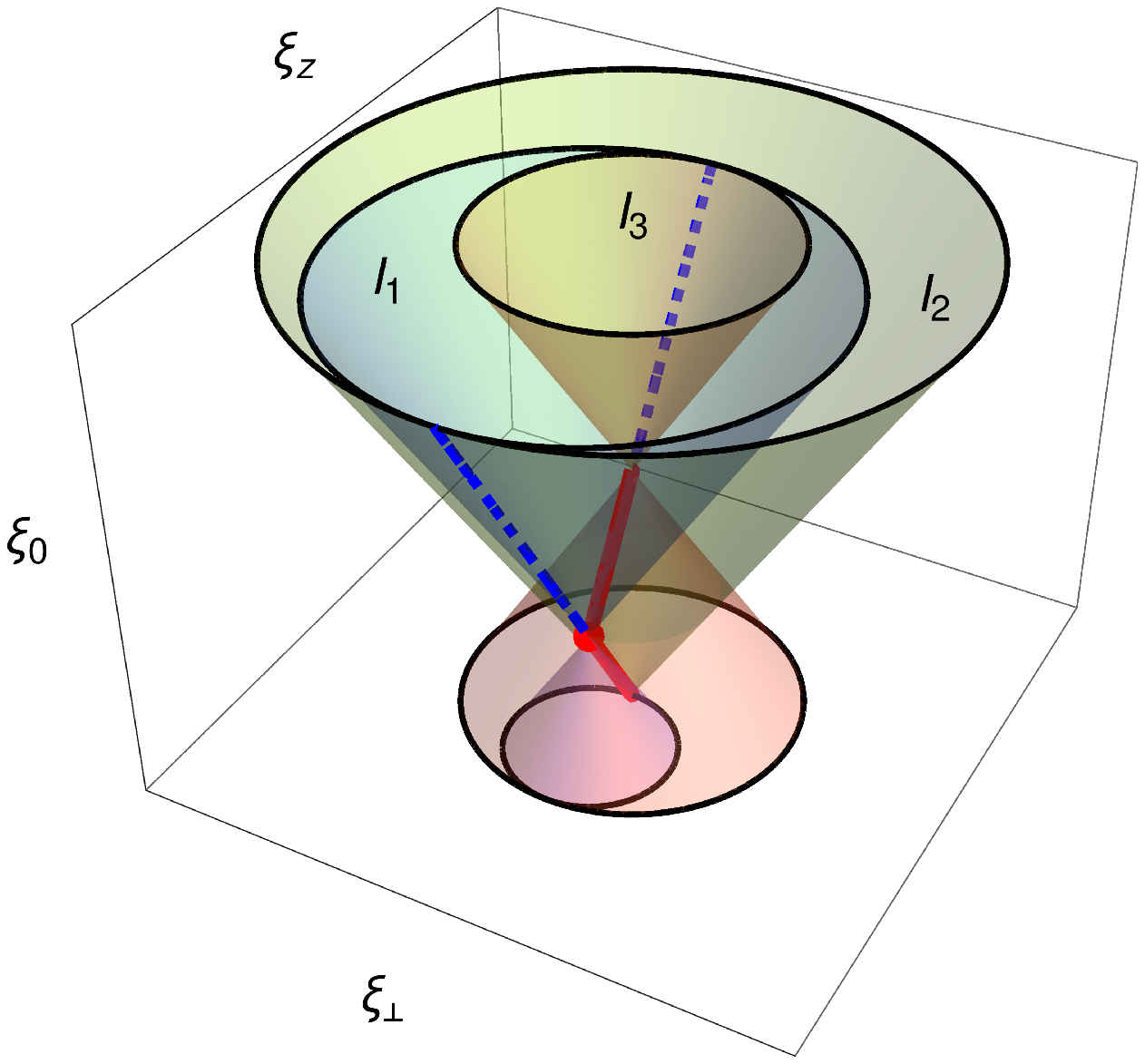}
\includegraphics[width=6cm]{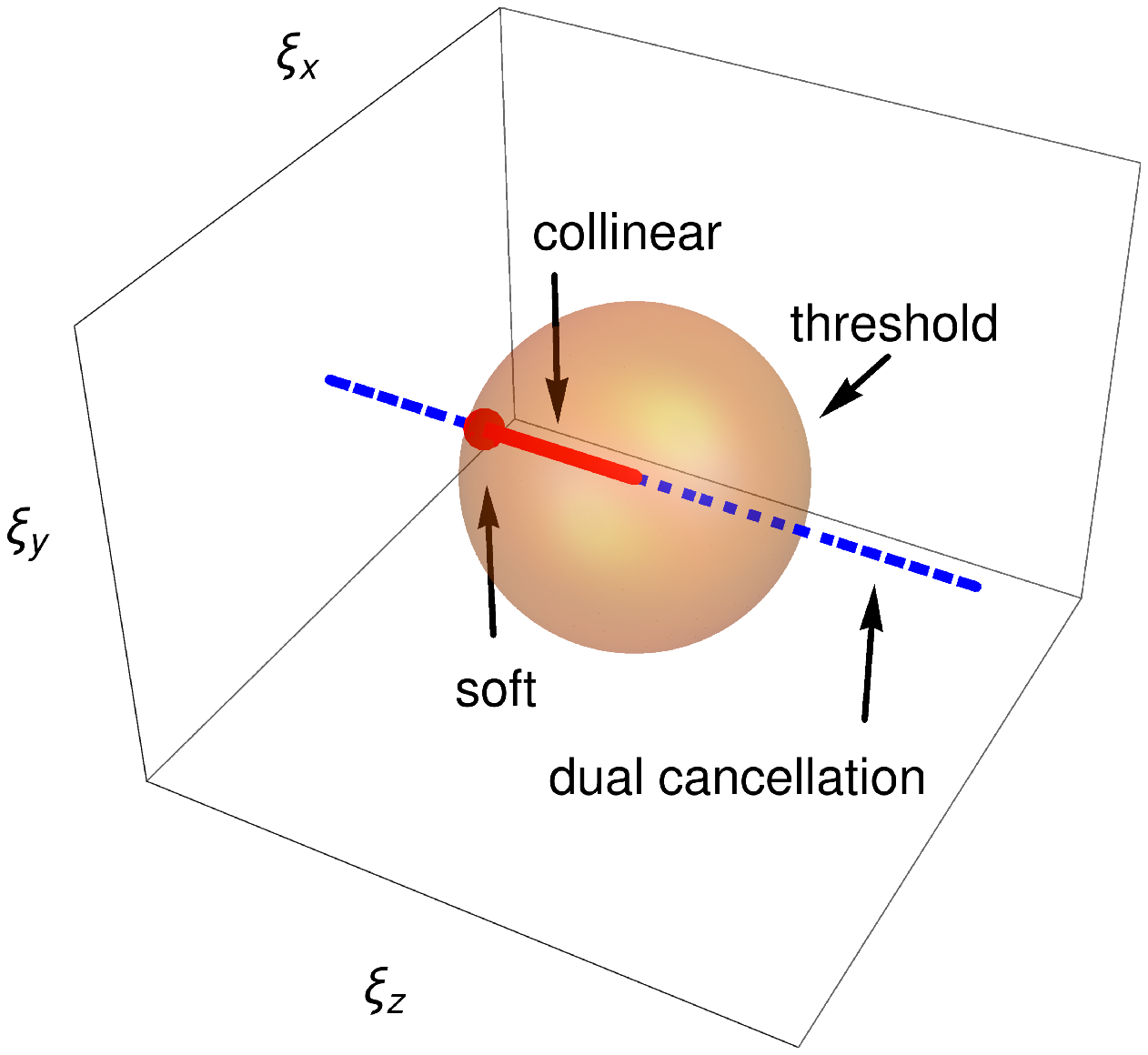}
\caption{Light-cones of the massless three-point function in the loop 
coordinates $\ell = \sqrt{s_{12}}/2 \, (\xi_0, \xi_\perp, \xi_z)$, 
with $\xi_\perp=\sqrt{\xi_x^2+\xi_y^2}$ (left). 
Each forward light-cone correspond to the integration region 
of one dual integral $I_i$, and the intersection with the other light-cones 
generates the singularities of the integrand. Location of threshold and IR singularities in the loop 
three-momentum space (right). 
\label{fig:cartesean}}
\end{center}
\end{figure}

To validate the LTD method with an explicit example, 
we focus our attention to the simplest case of the scalar three-point 
function, with $p_3\to p_1+p_2$ as external momenta and $p_3^2=s_{12}>0$; $p_1$, $p_2$ and the internal momenta are taken massless.
The loop integral is defined by the internal momenta $q_1 = \ell + p_1$,  
$q_2 = \ell + p_{12}$ and $q_3 = \ell$, according to Fig.~\ref{fig:feynman}.
We proceed step by step, by integrating first in DREG
and then by repeating the calculation in a suitable representation where 
the limit $\ep \to 0$ can safely be taken at the integrand level. 
The dual representation of the scalar three-point function consists of three 
contributions: 
\beq
I_i = - \int_\ell \td{q_i} \prod_{j\ne i} G_D(q_i; q_j)~,  \qquad i,j =1,2,3~.
\eeq
Assuming $p_1$ oriented along the positive $z$-axis, $p_2$ in the 
opposite direction, the on-shell loop momenta are parametrized as 
\beq
q_i^\mu = \frac{\sqrt{s_{12}}}{2} \, \xi_{i,0} \, 
(1, 2\sqrt{v_i(1-v_i)} \, {\bf e}_{i,\perp}, 1-2v_i)~,
\label{eq:qparam}
\eeq
with $\xi_{i,0} \in [0,\infty)$, $v_i \in [0,1]$, and ${\bf e}_{i,\perp}$ 
a unit vector in the transverse space.
Taking into account these parametrizations the scalar products are given 
by $2 q_i\cdot p_1 / s_{12} = \xi_{i,0}\, v_i$,
and $2 q_i\cdot p_2 / s_{12} = \xi_{i,0}\, (1-v_i)$, 
and the dual integrals are written as
\bea
I_1 &=& \frac{1}{s_{12}} \, \int d[\xi_{1,0}] \, d[v_1] \,  
\xi_{1,0}^{-1} \, (v_1 (1-v_1))^{-1}~, \nn \\
I_2 &=& \frac{1}{s_{12}} \, \int d[\xi_{2,0}] \, d[v_2] \,  
\frac{(1-v_2)^{-1}}{1 - \xi_{2,0} + i0}~, \nn \\
I_3 &=& - \frac{1}{s_{12}} \, \int d[\xi_{3,0}] \, d[v_3] \,  
\frac{v_3^{-1}}{1 + \xi_{3,0}}~,
\label{eq:integrals}
\eea
where we defined the $d$-dimensional integration measures as 
\bea
&& d[\xi_{i,0}] = \frac{\mu^{2\ep} \, (4\pi)^{\ep-2}}{\Gamma(1-\ep)} \,
s_{12}^{-\ep} \, \xi_{i,0}^{-2\ep} \, d\xi_{i,0}~, \nn \\
&& d[v_i] = (v_i(1-v_i))^{-\ep} \, dv_i~,
\eea
and the dual vector was taken as $\eta^\mu = (1,{\bf 0})$.
Figure~\ref{fig:cartesean}~(left) shows a graphical representation 
of the corresponding light-cones and their intersections 
where threshold and IR singularities appear.
In~\Eq{eq:integrals}, only the dual prescription of the dual 
integral $I_2$ has been kept explicitly, as it regulates a 
threshold singularity. The dual prescription of the other propagators
is required for the consistent partial cancellation of singularities 
among dual integrals~\cite{Buchta:2014dfa}, but can be removed
for this practical calculation.

The explicit computation in DREG of the total 
integrals in~\Eq{eq:integrals} gives
\beq
I_1 = 0~, \qquad
I_2 = \cgt \, \frac{\mu^{2\ep}}{2 \ep^2} \, s_{12}^{-1-\ep}\, e^{i2\pi \ep}~, \qquad
I_3 = \cgt \, \frac{\mu^{2\ep}}{2 \ep^2} \, s_{12}^{-1-\ep}~,  
\label{eq:i2i3}
\eeq
with
\beq
\cgt = c_\Gamma \, \frac{1}{\cos(\pi \epsilon)}~, \qquad
c_\Gamma = \frac{\Gamma(1+\epsilon)\Gamma^2(1-\epsilon)}
{(4\pi)^{2-\epsilon}\Gamma(1-2\epsilon)}~, 
\eeq
where $\cgt$ is the volume factor of the phase-space integrals, and 
$c_\Gamma$ is the volume factor of the loop integrals. The dual integral $I_1$ vanishes because it is scaleless, 
and following the usual convention in DREG it is analytically continued 
by matching its UV and IR poles, i.e. $\epsilon_{\uv}=\epsilon_{\rm IR}$. Summing up the results
in \Eq{eq:i2i3} we obviously obtain the well-known result 
\beq
L^{(1)}(p_1, p_2, -p_3) 
 = \int_\ell  \prod_{i=1}^3 G_F(q_i) 
 =  - c_\Gamma \, \frac{\mu^{2\ep}}{\ep^2} \, (-s_{12}-i0)^{-1-\ep}~. 
\label{eq:triangle}
\eeq

At this point, we start the discussion by isolating the singular regions of the 
loop integrand, which are bounded to a compact region of the loop 
three-momentum, as it can be seen from Fig.~\ref{fig:cartesean}~(right). The dual integral $I_1$ develops a soft singularity at $\xi_{1,0}=0$
and a collinear singularity at $v_1=0$ with $\xi_{1,0} \in [0,1]$, namely $\xi_z\in [-1,0]$;
the collinear singularity at $\xi_{z} > 0$ cancels with the other dual contributions. Likewise, the dual integral $I_2$ is collinear at $v_2=1$ with 
$\xi_{2,0} \in [0,1]$, or $\xi_z \in [-1,0]$, and undergo cancelling collinear 
divergences at $\xi_{z} < -1$. The integrand of $I_2$ gets in addition 
a threshold singularity at $\xi_{2,0} = 1$, which prevents us from having 
a well defined integral if we restrict the integration region by 
$\xi_{2,0} = 1$. Consequently, we introduce an energy-cut $w$, 
with $w$ close to zero, and define the following soft and collinear contributions 
to the dual integrals
\bea
\label{eq:softcoll0}
I_1^{\rm (s)} &=& I_1(\xi_{1,0}\le w)~, \nn \\  
I_1^{\rm (c)} &=& I_1(w \le \xi_{1,0} \le 1~; v_1\le1/2)~, \nn \\
I_2^{\rm (c)} &=& I_2(\xi_{2,0} \le 1 + w~; v_2\ge1/2)~.
\eea
The soft integral $I_1^{\rm (s)}$ contains the soft singularity of the 
dual integral $I_1$, and its collinear singularities close to the 
soft region, $\xi_{1,0} < w$. Once the soft part is removed, the 
remaining collinear singularities of $I_1$ are contained in the 
collinear integral $I_1^{\rm (c)}$. It is necessary to extend 
the integration limits of the collinear integral $I_2^{\rm (c)}$ 
beyond the strict collinear singular region of $I_2$ in order to 
treat properly its threshold singularity at $\xi_{2,0} = 1$.
Integrating these contributions in DREG,
we obtain the following results:
\bea
\label{eq:softcoll2}
I_1^{\rm (s)} &=& \cgt \, \frac{w^{-2\ep}}{\ep^2} \, \mu^{2\ep} 
\, s_{12}^{-1-\ep} \, \frac{\sin (2\pi \ep)}{2\pi \ep}~, \\  
I_1^{\rm (c)} &=& 
\cgt \, \frac{1 - w^{-2\ep}}{2 \ep^2} \, \mu^{2\ep} 
\, s_{12}^{-1-\ep} \, \frac{\sin (2\pi \ep)}{2\pi \ep}~, \nn \\
I_2^{\rm (c)} &=& 
\cgt \, \frac{\mu^{2\ep}}{4\ep^2} \, s_{12}^{-1-\ep} \,
\left(1+\frac{4^{\ep} \Gamma(1-2\ep)}{\Gamma^2(1-\ep)} \right) \, 
\left[ e^{i 2 \pi \ep} - w^{-2\ep} \, 
{}_2F_1\left(2\ep, 2\ep, 1+2\ep; - \frac{1}{w}\right) 
\, \frac{\sin (2\pi \ep)}{2\pi \ep} \right]~, \nn
\eea 
with ${}_2F_1\left(2\ep, 2\ep, 1+2\ep; z\right) = 1+ 4\ep^2 \, {\rm Li}_2(z)
+ {\cal O}(\ep^3)$. As expected, the soft integral $I_1^{\rm (s)}$ contains double poles in $\ep$, 
while the collinear integrals $I_1^{\rm (c)}$ and $I_2^{\rm (c)}$
develop single poles only. Each integral depends on the cut $w$, but  
the IR poles of the sum are independent of $w$ and they agree with 
the total divergences of the full integral in~\Eq{eq:triangle}.

At this stage, we have shown for the first time with a concrete example
that the IR singularities in the LTD representation
are constrained to a compact region of the loop three-momentum, 
as it was anticipated before based on the general demonstration from  
Ref.~\cite{Buchta:2014dfa}, and have recalculated 
the IR poles only by taking into account this compact region. 
The definition of the soft and collinear integrals is, however, 
somehow arbitrary; strictly, the IR singularities of the loop integrand 
are constrained to the region defined by the limit $w\to 0$ 
and more precisely to the segment defined by $\xi_z \in [-1,0]$.
Any compact region of the loop three-momentum enclosing $\xi_z \in [-1,0]$
will lead to the same $\epsilon$ poles, although to different finite 
contributions. We will later redefine the IR regions of the 
loop integrand in a more convenient way to combine them with the real corrections.  

Outside the soft and collinear regions, the sum of the dual integrals is finite because
the integrand singularities appear at the intersection of forward 
light-cones. We define first a forward region with $v_i\le 1/2$: 
\bea
I^{(\f)} &=& I_1(\xi_{1,0}\ge 1~; v_1\le\frac{1}{2}) + I_2(v_2\le\frac{1}{2})
+ I_3(v_3\le\frac{1}{2}) \nn \\
&=& c_\Gamma \, \frac{1}{s_{12}}
\int_{0}^\infty d\xi_{0} \, \int_0^{1/2} dv \, \bigg[ 
\frac{1}{1 + \xi_{0}} \bigg( (1-v)^{-1} \nn \\ &+& 
2 \, \ln{\frac{1+\xi_0}{\xi_{0}}} \, \delta(v) \bigg) 
+ \frac{(1-v)^{-1}}{1-\xi_0 + i0} \bigg] + {\cal O}(\ep)~. 
\label{eq:forward}
\eea
The dual integrals $I_1$ and $I_3$ are divergent at $v_1=v_3=0$ in the forward region,
but their sum is finite in this collinear limit. However, they diverge in 
the UV in all directions in spite of the absence of UV divergences 
in the original integral. This is due to the fact that 
dual propagators are linear in the loop momentum. The UV finiteness 
of the original integral is recovered by adding together all 
the dual integrals~\cite{Catani:2008xa}. Therefore, to cancel this UV behaviour 
it is necessary to add the contribution of $I_2$. The integral over the sum of the three dual integrands 
in~\Eq{eq:forward} can thus be performed with $\ep=0$. In this expression, we have also identified 
all the angular and energy variables, $v_i=v$, 
$\xi_{1,0}-1 = \xi_{2,0} = \xi_{3,0} = \xi_0$, 
to obtain common integration limits.
Notice that it is necessary to keep the dual $i0$ prescription 
of $I_2$ because its threshold singularity at $\xi_{2,0}=1$ 
appears within the integration region; it does not affect the UV cancellation. The logarithmic term of the integrand in~\Eq{eq:forward} 
is the result of the mismatch in the $\ep$-expansion of 
the integration measure of $I_1$ and $I_3$.
The same result is obtained without extra logarithmic
contributions by expressing ($\xi_{1,0},v_1$) in terms of 
($\xi_{3,0},v_3$) at the price, however, of 
introducing cumbersome integration limits. We obtain
in both cases 
\beq
I^{(\f)} = c_\Gamma \, \frac{1}{s_{12}} \, 
\left[ \frac{\pi^2}{3} - i\pi \, \log(2) \right] + {\cal O}(\ep)~.
\label{eq:forward2}
\eeq
The other finite contribution in the backward 
region ($v_i \ge 1/2$) is
\bea
I^{(\b)} &=&  I_1(\xi_{1,0} \ge w~; v_1\ge\frac{1}{2}) +
I_2(\xi_{2,0} \ge 1+w~; v_2\ge\frac{1}{2}) + I_3 (v_3\ge\frac{1}{2}) 
\nn \\ &=& c_\Gamma \, \frac{1}{s_{12}} \, \bigg[
2 \, {\rm Li}_2\left(-\frac{1}{w}\right) - \log(2) \log(w) \bigg]
+ {\cal O}(\ep)~. 
\label{eq:backward2}
\eea
Now, the collinear cancellation occurs at $v_1=1=v_2$ and 
the UV behaviour of $I_1+I_2$ is cancelled by $I_3$.
As expected, the sum of the soft, collinear, forward and backward 
integrals in \Eq{eq:softcoll2}, \Eq{eq:forward2} and \Eq{eq:backward2}
leads to the full result given in \Eq{eq:triangle}, up to ${\cal O}(\ep)$. 

\section{Real-Virtual mapping and cancellation of infrared singularities}
\label{sec:realvirtual}
In the previous section we have analyzed the divergent structure 
of a scalar one-loop integral, and have isolated its IR behaviour 
to a compact region of the loop three-momentum. We shall illustrate now 
how to combine virtual corrections with real radiation 
in order to obtain physical cross-sections. To show the advantages of 
the method it is enough to work out a simplified calculation where the one-loop 
scattering amplitude is just given by the scalar three-point function
\beq
| {\cal M}^{(1)} (p_1, p_2; p_3)\ra = 
-i \, g^3 \, s_{12}\, L^{(1)}(p_1, p_2, -p_3)~, 
\eeq
where $g$ is an arbitrary coupling. Explicit computations in a real 
physical model will be presented elsewhere \cite{Rodrigo:2015PREP,Sborlini:2016fcj}.
The corresponding scalar tree-level 
amplitude is $|{\cal M}^{(0)}(p_1, p_2; p_3) \ra = i g$, and 
we shall take into account their interference, 
${\rm Re}\, \la{\cal M}^{(0)}|{\cal M}^{(1)}\ra$, 
integrated over the two-body phase-space.   
We consider also scalar scattering amplitudes with emission 
of an extra particle $|{\cal M}^{(0)}_{ir} (p_1', p_2', p_r'; p_3)\ra = - i g^2 \sqrt{s_{12}}/ s_{ir}'~,$
and in particular the interference (see Fig.~\ref{fig:feynman}~(right))
\beq
{\rm Re} \, \la{\cal M}^{(0)}_{2r} |{\cal M}^{(0)}_{1r}\ra = 
g^4 \, \frac{s_{12}}{s_{1r}'\, s_{2r}'}~,
\eeq
which is integrated over the corresponding three-body phase-space~\footnote{For 
identical final-state particles all the other possible interferences should 
also be considered. Besides, the squared amplitudes $|{\cal M}_{ir}|^2$ are necessary, 
which have to be mapped to self-energy insertions of external legs. 
For the current illustrative purposes of the paper, however, it is enough to consider 
this interference.}.
To combine virtual and real corrections directly, we first decompose the real-emission phase-space by using the identity 
$1 = \theta(y_{2r}'-y_{1r}') + \theta(y_{1r}'-y_{2r}')$, where 
$y_{ir}'= s_{ir}'/s_{12}$. It is unnecessary to single out a soft region. 
In general, segmenting the real-emission phase-space 
into different regions featuring at most one soft and/or collinear singularity 
is a common practice in many subtraction schemes, as for 
instance in FKS~\cite{Frixione:1995ms}. The main novelty of the LTD approach
is that that segmentation will be translated into the virtual contributions such 
that a local cancellation of singularities at the integrand level is achieved. 

Thus, we define the following 
dual contributions to the total cross-section 
\bea
\widetilde \sigma_{\r,i}^{(1)} &=& \frac{1}{2 s_{12}} \,  \int d\Phi_{1\to 3} \, 
2 {\rm Re} \, \la{\cal M}^{(0)}_{2r} |{\cal M}^{(0)}_{1r}\ra \, \theta(y_{jr}'-y_{ir}')~,  
\nn \\
\widetilde \sigma_{\v,i}^{(1)} &=& \frac{1}{2 s_{12}} \, \int d\Phi_{1\to 2} \, 
2 {\rm Re} \, \la{\cal M}^{(0)} |{\cal M}_i^{(1)}\ra \, \theta(y_{jr}'-y_{ir}')~,  
\qquad i,j = 1, 2~,
\label{eq:sigmas}
\eea
where $|{\cal M}_i^{(1)}\ra = - i g^3 \, s_{12} \, I_i$ is 
the $i^{\rm th}$ dual component of the one-loop scattering amplitude.
The third dual component given by $I_3$ in~\Eq{eq:integrals} does not contribute to the 
IR structure of $|{\cal M}^{(1)}\ra$, and can be integrated independently of the real 
corrections.

The overall picture presented in~\Eq{eq:sigmas} must be completed 
with a mapping of the virtual and the real kinematics.
For a fixed value of the external momenta $p_1$ and $p_2$ 
entering the loop scattering amplitude, 
and the loop three-momentum $\boldsymbol{\ell}$, 
we define the following mapping, with $q_1=\ell+p_1$ on-shell,
\bea
&& p_r'^{\mu} = q_1^\mu~, \qquad
p_1'^{\mu} = - q_{3}^\mu + \alpha_1 \, p_2^\mu 
= p_{1}^\mu - q_{1}^\mu + \alpha_1 \, p_2^\mu~,  \nn \\
&& p_2'^{\mu} = (1-\alpha_1) \, p_2^{\mu}~,  \qquad 
\alpha_1 = \frac{q_3^2}{2q_3\cdot p_2}~, 
\label{eq:mapping1} 
\eea
where $q_3 = q_1-p_1$ is off-shell. 
Momentum conservation is automatically 
fulfilled as $p_1'+p_2'+p_r'=p_1+p_2$, and all the primed
final-state momenta are on-shell. The non-collinear parton 
$p_2'$ plays here the role of the spectator parton in the dipole 
formalism~\cite{Catani:1996vz}. The mapping in~\Eq{eq:mapping1} is also valid for an arbitrary number 
of external particles, provided that $p_i'=p_i$ for partons
not entering in the collinear subprocess. 
The two-body invariants are given by
\beq
y_{1r}' = \frac{v_1 \, \xi_{1,0}}{1-(1-v_1) \, \xi_{1,0}}~, \qquad 
y_{2r}' = \frac{(1-v_1) \, (1-\xi_{1,0}) \, \xi_{1,0}}{1-(1-v_1) \, \xi_{1,0}}~, \qquad
y_{12}' = 1 - \xi_{1,0}~.
\label{eq:twobody}
\eeq
These expressions are obtained from the momentum parametrizations 
given in~\Eq{eq:qparam} and~\Eq{eq:mapping1}.
The loop momentum $q_1$ and $p_1$ become collinear at $v_1\to 0$, 
with $\alpha_1 \to 0$ and $y_{1r}'\to 0$ in this limit. 
According to \Eq{eq:sigmas}, we use this mapping in the regions 
of the two-body and three-body phase-space where $y_{1r}'< y_{2r}'$: 
\bea
\widetilde  \sigma_1^{(1)} &=& \widetilde \sigma_{\v,1}^{(1)} + \widetilde \sigma_{\r,1}^{(1)} 
= \sigma^{(0)} \, 2 g^2 \, \int d[\xi_{1,0}] \, d[v_1] \, \theta(1-2 v_1) \,
\theta\left(\frac{1-2v_1}{1-v_1}-\xi_{1,0}\right) \nn \\ &\times&
\xi_{1,0}^{-1} \, (v_1 (1-v_1))^{-1} \, \left[  
\left(\frac{1-\xi_{1,0}}{1-(1-v_1)\, \xi_{1,0}} \right)^{-2\ep} - 1 \right]~, 
\label{eq:sigma1}
\eea
where $\sigma^{(0)} = g^2/(2 s_{12}) \int d\Phi_{1\to 2}$ would be the 
lowest order cross-section. In~\Eq{eq:sigma1}, the three-body phase-space 
$d\Phi_{1\to 3}$ was written in terms of the loop variables $\xi_{1,0}$ and $v_1$
by using~\Eq{eq:twobody}. The integrand in \Eq{eq:sigma1} has the 
form $\xi_{1,0}^{-1-2\ep} \, v_1^{-1-\ep} \, f(v_1,\xi_{1,0})$, 
but $f(v_1=0) = 0 = f(\xi_{1,0}=0)$ with $f(v_1,\xi_{1,0}) = {\cal O}(\ep)$, 
and thus this integral is of 
order $\ep$, i.e. $\widetilde \sigma_1^{(1)} = {\cal O} (\ep)$. 
Similarly, we define a complementary mapping to cover the remaining 
three-body phase-space, which includes the other collinear
limit with $y_{2r}'\to 0$. Considering $q_2$
on-shell, the corresponding mapping is
\bea
&& p_2'^{\mu} = q_2^\mu~, \qquad
p_r'^{\mu} = - q_{1}^\mu + \alpha_2 \, p_1^\mu 
= p_{2}^\mu - q_{2}^\mu + \alpha_2 \, p_1^\mu~,  \nn \\
&& p_1'^{\mu} = (1-\alpha_2) \, p_1^{\mu}~,  \qquad 
\alpha_2 = \frac{q_1^2}{2q_1\cdot p_1}~, 
\label{eq:mapping2} 
\eea 
and the associated invariants are 
\beq
y_{12}' = \frac{v_2 \, (1-\xi_{2,0}) \, \xi_{2,0}}{1- v_2 \, \xi_{2,0}}~, \qquad 
y_{2r}' = \frac{(1-v_2) \, \xi_{2,0}}{1-v_2 \, \xi_{2,0}}~, \qquad
y_{1r}' = 1 - \xi_{2,0}~.
\eeq
Then,   
\bea
\widetilde \sigma_2^{(1)} = \widetilde \sigma_{\v,2}^{(1)} + \widetilde \sigma_{\r,2}^{(1)}  &=&
\sigma^{(0)} \, 2 g^2 \, \int d[\xi_{2,0}] \, d[v_2] \, 
\theta \left(\frac{1-\sqrt{1-v_2}}{v_2} - \xi_{2,0}\right)
\, (1-v_2)^{-1} \, \nn \\ &\times&
\bigg[\frac{(1-\xi_{2,0})^{-2\ep}}{(1-v_2\, \xi_{2,0})^{1-2\ep}} 
- \frac{1}{1-\xi_{2,0}+i0} - i \pi \delta(1-\xi_{2,0})\bigg]~, 
\label{eq:sigma2vr}
\eea
and we obtain directly with $\ep=0$
\beq
\widetilde \sigma_2^{(1)} = - \sigma^{(0)} \, a \, \frac{\pi^2}{6}
+ {\cal O}(\ep)~,
\label{eq:sigma2}
\eeq 
where $a=g^2/(4\pi)^2$. The condition $y_{2r}'<y_{1r}'$ excludes the threshold singularity of $I_2$ 
from the integration region with the exception of the single point at 
$v_2=1$, $\xi_{2.0}=1$. This makes unnecessary the introduction of an energy-cut $w$ as in Sec.~\ref{sec:cancel}.

Finally, analogously to the forward and backward dual integrals 
defined in Sec.~\ref{sec:cancel}, we collect the remnants of the 
three dual integrals obtained after excluding the respective regions of the loop 
three-momentum already included in~\Eq{eq:sigma1} and~\Eq{eq:sigma2}.
Explicitly, we define
\bea
\nn \bar \sigma^{(1)}_{\v} &=& \sigma^{(0)} \, 2 g^2 \, \left[ - \int d[\xi_{1,0}] \, d[v_1] \, 
\, \left( 1- \theta(1-2 v_1) \,
\theta\left(\frac{1-2v_1}{1-v_1}-\xi_{1,0}\right) \right) 
\xi_{1,0}^{-1} \, (v_1(1-v_1))^{-1} \right.
\\ \nn &-& \left. \int d[\xi_{2,0}] \, d[v_2] \, 
\theta \left(\xi_{2,0}-\frac{1-\sqrt{1-v_2}}{v_2}\right) \, (1-v_2)^{-1} \left(\frac{1}{1-\xi_{2,0}+i0} + i \pi \delta(1-\xi_{2,0})\right) \right. 
\\ &+& \left. \int d[\xi_{3,0}] \, d[v_3] \, 
\vphantom{\theta \left(\xi_{1,0}-\frac{1-2 v_1}{v_1}\right)} \, \frac{v_3^{-1}}{1+\xi_{3,0}} \right]~.
\label{eq:sigmaremexplicado} 
\eea
Although the individual contributions in~\Eq{eq:sigmaremexplicado} are singular, 
this virtual cross-section is UV and IR finite, and can be calculated 
with $\ep = 0$. We apply the change of variables used to 
obtain~\Eq{eq:forward2} and~\Eq{eq:backward2} to avoid logarithmic contributions
from the $\ep$-expansion of the integration measure. The result is 
\beq
\bar \sigma^{(1)}_{\v} 
= \sigma^{(0)} \, a \, \frac{\pi^2}{6} + {\cal O}(\ep)~.
\label{eq:sigmafb}
\eeq
The sum of all the contributions, \Eq{eq:sigma1}, \Eq{eq:sigma2} and~\Eq{eq:sigmafb},
gives a total cross-section of ${\cal O} (\ep)$, in agreement with 
the result that would be obtained from the standard calculation in DREG. 

To conclude this section, we want to emphasize that DREG was
only used for a consistency check of our results. The advantage of the method presented 
here is that NLO cross-sections can be computed in four dimensions, i.e. with $\ep =0$, 
because soft and collinear divergences of virtual and real corrections 
are matched locally at the integrand level. Collinear factorization 
guarantees that this matching is fulfilled in QCD~\cite{Buchta:2014dfa,Catani:2011st}.
Threshold singularities should be treated apart in numerical calculations, 
e.g. by contour deformation~\cite{Soper:1998ye,Becker:2010ng}.
A first numerical implementation of multi-leg scalar and tensor one-loop integrals 
in LTD has been presented in Ref.~\cite{Buchta:phd,Buchta:2015vha,Buchta:2015wna}.

\section{Ultraviolet Renormalization}
\label{sec:UV}
LTD also offers an appealing physical interpretation 
for the renormalization of UV divergences 
at the integrand level. To illustrate the cancellation of these divergences, 
we consider the simplest purely UV scalar two-point function, with massless internal lines. 
A detailed calculation of this scalar two-point function in the LTD framework has been presented in Ref.~\cite{Catani:2008xa}. 
Following the suggestion of Ref.~\cite{Becker:2010ng}, 
we construct its UV counter-term from
\beq
I_{\uv}^{\cnt} = \int_\ell \frac{1}{\left(q_{\uv}^2-\mu_{\uv}^2+i0\right)^2}~,
\label{uvint}
\eeq
where $q_{\uv} = \ell + k_{\uv}$, with $k_{\uv}$ located somewhere 
in the loop momentum space. This counter-term has the same local UV behaviour as 
the integrand of the scalar two-point function in all the loop directions.
We shall now find the dual representation of $I_{\uv}^{\cnt}$ in three dimensions. 

\begin{figure}[ht]
\begin{center}
\includegraphics[width=6cm]{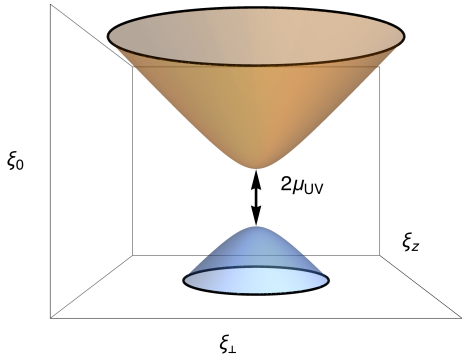}
\caption{On-shell hyperboloids of the ultraviolet counter-term.  
\label{fig:carteseanUV}}
\end{center}
\end{figure}

The dual representation of~\Eq{uvint} requires to deal with double 
poles in the loop energy~\cite{Bierenbaum:2012th}, 
with the UV propagator being massive. The calculation of the corresponding 
residue leads to the dual representation 
\beq
I_{\uv}^{\cnt} = \int_{\ell} \, \frac{\td{q_{\uv}}}{2\, \left(q_{\uv,0}^{(+)}\right)^2}~,
\eeq
where $q_{\uv,0}^{(+)} = \sqrt{\qb_{\uv}^2 + \mu_{\uv}^2 - i0}$. The detailed calculation of the double pole residue 
has been presented in Ref.~\cite{Bierenbaum:2012th}; 
again we used $\eta^\mu = (1,{\bf 0})$ as dual vector. The arbitrary scale $\mu_{\uv}$ admits a direct physical interpretation 
as renormalization scale because the UV counter-term only contributes 
for loop energies larger than $-k_{\uv,0}+\mu_{\uv}$ in the LTD representation (see Fig.~\ref{fig:carteseanUV}), 
although with unconstrained loop three-momentum.  
As renormalization scale, $\mu_{\uv}$ has to be chosen of the order
of the physical hard scale. In addition, a clear criteria to select 
$\mu_{\uv}$ and $k_{\uv}$ arises if we pretend that the on-shell 
hyperboloids of the UV propagator in~\Eq{uvint} do not intersect 
with any of the forward on-shell hyperboloids of the original integral. 
Since the distance between the UV forward and backward on-shell hyperboloids 
is $2\mu_{\uv}$, the minimal choice fulfilling the aforementioned conditions 
would be $1/2$ of the hard scale, which accidentally agrees with one of the standard 
limits used to estimate the theoretical uncertainty. Of course, this is an 
\emph{ad hoc} argument that is not supported by any physical evidence, 
but it represents a good interpretation of that choice.

In consequence, subtracting the UV counter-term from the scalar two-point function we obtain 
\beq
L^{(1)}(p,-p) - I_{\uv}^{\cnt} = \frac{1}{(4\pi)^2} \, 
\left[ - \log\left( - \frac{p^2}{\mu_{\uv}^2} -i0 \right)+ 2 \right]
+ {\cal O}(\ep)~,
\label{uvfinite}
\eeq
and this calculation gives the same result if each of the individual contributions 
is integrated in DREG or if the sum of the integrands is taken at $\ep=0$.

We conclude this section by noting that 
contrary to the four-dimensional UV renormalization introduced in Ref.~\cite{Pittau:2012zd}, 
where the limit $\mu \to 0$ is taken with the purpose of removing the polynomial dependence 
in $\mu$ and the remainder is evaluated at the renormalization scale $\mu=\mu_{\rm R}$,
the counter-term defined by~\Eq{uvint} keeps the complete dependence in $\mu_{\uv}$.
Eventual polynomial dependences in $\mu_{\uv}$ can be cancelled by adding to the counter-term 
UV subleading contributions such that only the UV pole is subtracted~\cite{Becker:2010ng}.  

\section{Conclusions}
\label{sec:conclusions}

We have applied the LTD formalism to physically reinterpret perturbative calculations in QFT.
With some reference examples, we have reanalyzed the infrared and causal 
structure of NLO computations. The sum over degenerate IR states is accomplished at integrand
level by mapping the phase-space of real radiation with that of the virtual contributions
in the region of the loop three-momentum which is of the size of the physical scales. 
This represents a new paradigm in perturbative calculations as virtual and real corrections 
are not regularized independently, but they act directly as the IR subtraction 
counter-term of each other. Outside this region,
at large loop three-momentum, virtual corrections require only the subtraction 
of UV divergences at integrand level to render them finite and renormalization 
acquires a clear physical interpretation. 

The application of LTD to loop scattering amplitudes in 
gauge theories was already discussed in Ref.~\cite{Catani:2008xa}, 
and the momentum mappings introduced in this paper can be extended straightforwardly to massless $n$-body scattering processes, and even generalized to the massive case. 
Each mapping is suitable for a different collinear configuration, 
and has to be applied to the sum of diagrams belonging to gauge invariant subsets 
featuring the same collinear limit. 

The results presented in this paper represent a promising first effort 
towards a four-dimensional implementation for the computation of physical 
cross-sections at NLO without introducing soft and final-state 
collinear subtractions. They also open the attractive possibility of extending 
this approach to next-to-next-to leading order (NNLO) and beyond.
Further details will be given in a forthcoming publication, including the 
application to gauge theories \cite{Rodrigo:2015PREP}.

\section*{Acknowledgements}

We thank S. Catani for a longstanding fruitful collaboration. 
This work has been supported by the Research Executive Agency (REA) 
under the Grant Agreement No. PITN-GA-2010-264564 (LHCPhenoNet), by CONICET Argentina,  
by the Spanish Government and ERDF funds from the European Commission (Grants
No. FPA2014-53631-C2-1-P, FPA2011-23778, 
CSD2007-00042 Consolider Project CPAN) and by Generalitat
Valenciana under Grant No. PROMETEOII/2013/007.
The work of RJHP is partially supported by CONACyT, M\'exico.

\end{document}